# Illegal Border Cross Detection and Warning System Using IR Sensor and Node MCU

Afsana Tasnim, Shawan Shurid, and AKM Bahalul Haque

***Abstract*—We often find illegal immigrants moving from one country to another. By means of land these illegal immigrants move over the fence cut the border wires and moves to the other part of the land. What do you think our soldiers are not doing their job? It is not that actually. It is very difficult to watch consistently over the border. Soldiers have a limited vision to check around the whole land mass. So for them we have come up with a solution that we have built a device that will sense the presence of an intruder (illegal immigrant). This device will be installed over the fences. When an intruder passes over the fence this device will transmit the signal to the soldiers Smartphone app (BLYNK app). The soldier will be notified with the signal and after receiving the signal, the soldier can switch on the alarm and the emergency lights via the app. By this the soldiers in the camp will be alerted and can take their respective positions and arrest the intruder which was passing the border illegally. By this device we can alert the soldiers in the border to take more safety precautions to keep our country safe.**

***Index Terms*—Node MCU ESP32, infrared sensor, Blynk app.**

## I. Introduction

### *A. Description*

In Villages and in the areas near the border of a country people and the citizen of our country literally face many problems regarding security issues. People of different tribes of the other side of the border do come inside the border and attack our own civilians. Mostly they do a lot of harassment and they dominate over this people. They come and steal expensive items from the civilians. An estimate made in the year 2000 placed the total number of illegal Bangladeshi immigrants in India at 1.5 crore, with around 3 lakhs entering every year. The rule of thumb for such illegal immigrants is that for each illegal person caught four get through. While many immigrants have settled in the border areas, some have moved on, even to faraway places such as Mumbai and Delhi. The trip to India from Bangladesh is one of the cheapest in the world, with a trip costing around Rs.2000 (around $30 US), which includes the fee for the "Tour Operator". As Bangladeshi are cultural similar to the Bengali people in India, they are able to pass off as Indian citizens and settle down in any part of India to establish a far better future than they could in Bangladesh, for a very small price. This false identity can be bolstered with false documentation available for as little as Rs.200 ($3 US) can even make them part of the vote bank. Although there is fancy or boundary for the safety purpose but it's not enough for security. The main purpose of the project is to enhance the border security electronically with automation and with that to reduce this kind of problems.

Afsana Tasnim, Shawan Shurid, and AKM Bahalul Haque are with North South University, Dhaka, Bangladesh (e-mail: {afsana.tasnim, shawan.shurid, bahalul.haque}@northsouth.edu).

### *B. Solution*

And Using this concept we can easily identify if any strangers entering the border. And by this system we can make our border soldiers more alert. We are here with a solution that will help them to stay more alert regarding the intruders entering over the border. We will provide a device which will be there across the pillars of the border to detect the presence of any intruder entering the country. If any intruder is found entering the system will alert the soldier to take action. And automatically a buzzer will turn on to alert the soldier base camps.

## II. Results Literature Review

Many analysts have been working on IOT and wireless sensor zones to provide the best security component. In this segment, we described different intrusion detection systems which are proposed in recent years. We have experienced different research papers and found what a were different security measures were used and what different strategy and thoughts overcome the problems.

### *A. Security Cameras*

Security cameras are of numerous sorts with obviously many element inclinations. Settling on the camera for use in a home setting can be an overwhelming errand given the various alternatives. Despite the fact that these cameras are accessible in a wide scope of sizes, fields of view, the nature of pictures, and distinctive scopes of movement, surveillance cameras utilized in homes all give video pictures of events inside a field of view. The cameras can show the moves making place continuously or record everything for later view. A portion of the camera frameworks empowers clients to watch and control their surveillance cameras on the web. In Fig. 1 shows a surveillance camera. Surveillance cameras are sorted into those utilized inside and outside. Every one of the classifications has diverse camera styles. Further separation, all around, isn't essential on the grounds that the variety is evident amid the assessment of highlights. Surveillance cameras can either come as remain solitary units with introduced applications for observing the framework or as a major aspect of a bundle which clients buy in to for arrangement of home security.

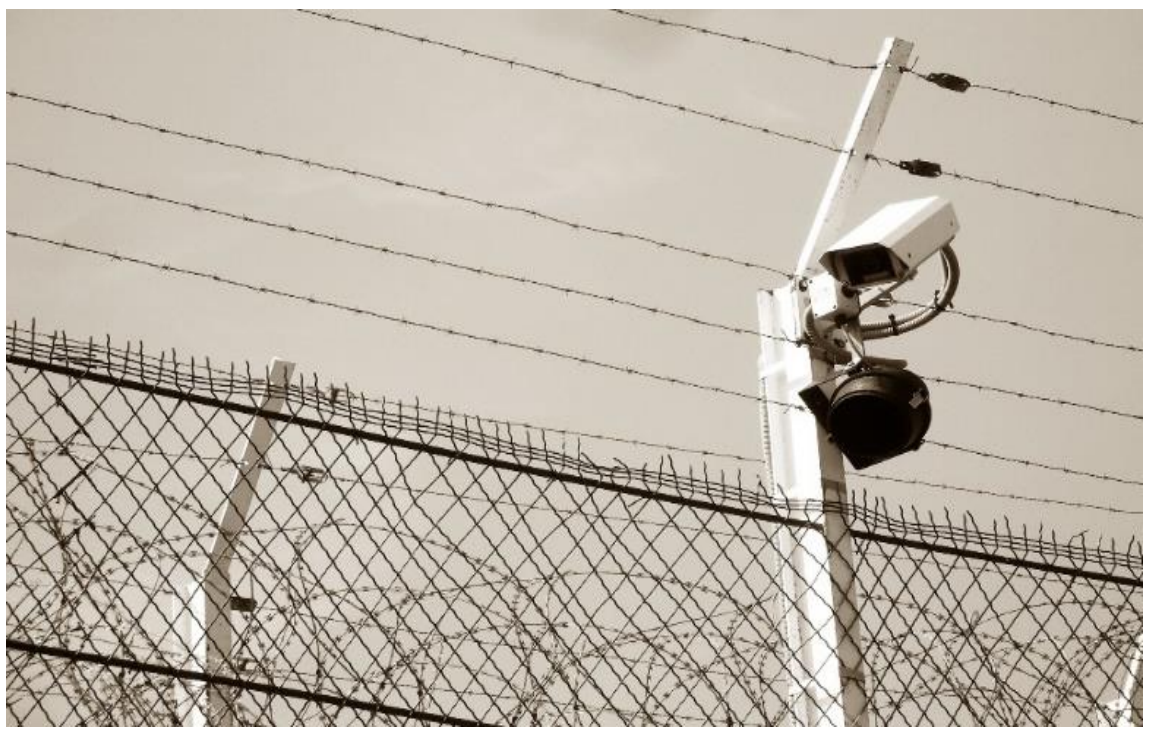

Fig. 1. Surveillance cameras in the border areas which continuously keeps monitoring.

### *B. Microwaves*

Ref. [1] Yaskawa and M. Kim to distinguish interlopers, there are strategies utilizing infrared sensors, camcorders, and microwave spatial qualities. Be that as it may, security encroachment and constrained identification extend are intrinsic issues in the infrared sensor and camcorder, individually. Notwithstanding when microwaves are utilized, the exactness breaks down except if high floods of intensity are impeded. Along these lines, in this exploration, gatecrasher recognition is performed by utilizing the spatiotemporal attributes of microwave multipath engendering. In this strategy, utilizing cluster beamforming at different defer taps improves the discovery likelihood.

### *C. GSM*

Ref. [2] R. Newlin Shebiah, B. Deeksha, and S. Aparna proposed a system for joining the classifier into the course is done which in the long run guarantees the establishment extraction and pictures only the object of ability. The proposed structure is attempted with the animal database and if the wild animals are recognized, by then the messages are sent through the GSM shown in Fig. 2. The camera interminably records the scene and once the development is found, by then it gets the photos and uses the portrayal methods to perceive its trademark features. As such, when the animals are recognized as risky using the request strategies, the writings are sent to the ranchers to safeguard themselves and their cultivating territories. The inconvenience of this method is that it can't be most suitable to cover a far-reaching zone of land and isn't likely attainable for the area and following the development of various or social occasions of animals.

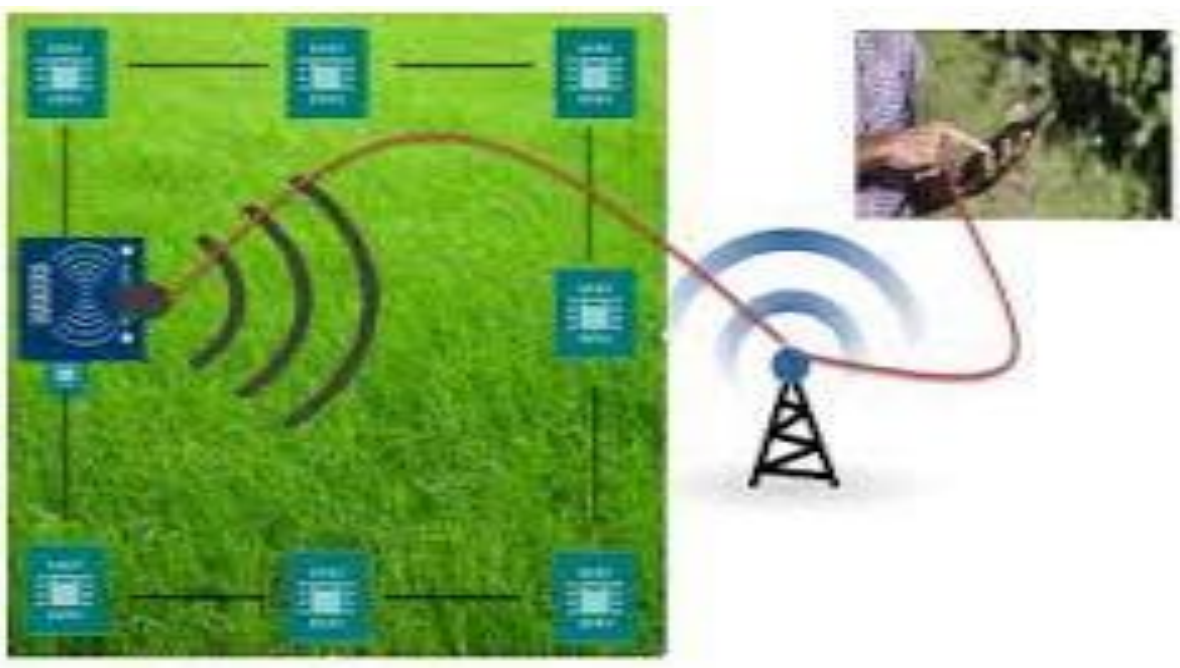

Fig. 2. Messages are sent through GSM to the farmers and the concerned people to safeguard themselves and their agricultural lands.

Ref. [3] Jyothsna and V. Prasad tells that Anomaly-based identification is based with respect to characterizing the system functioning. The system performance is as per the predefined network, at that point it is acknowledged or else it triggers the occasion in the irregularity identification. The acknowledged system conduct is arranged or learned by the details of the system managers.

### *D. Keystroke*

Ref. [4] There have been fixed approaches for login strategy like a Keystroke technique, titled “IMPROVED AUTHENTICATION MECHANISM USING KEYSTROKE ANALYSIS”. Yet, the disadvantage of that was that on the off chance that the user is mentally irritated, at that point the rhythm of his typing speed might differ and can cause a failed login. Key stroke analysis shown in Fig. 3. The last result of the system will help the owner with protecting his system and the personal information of the owner by not enabling the intruder to reach the system as the system can shut down. Every one of the ports can be blocked and can carry on. To carry out this we will use the GSM module which will tell the user at whatever point somebody tries to reach the owner 's system.

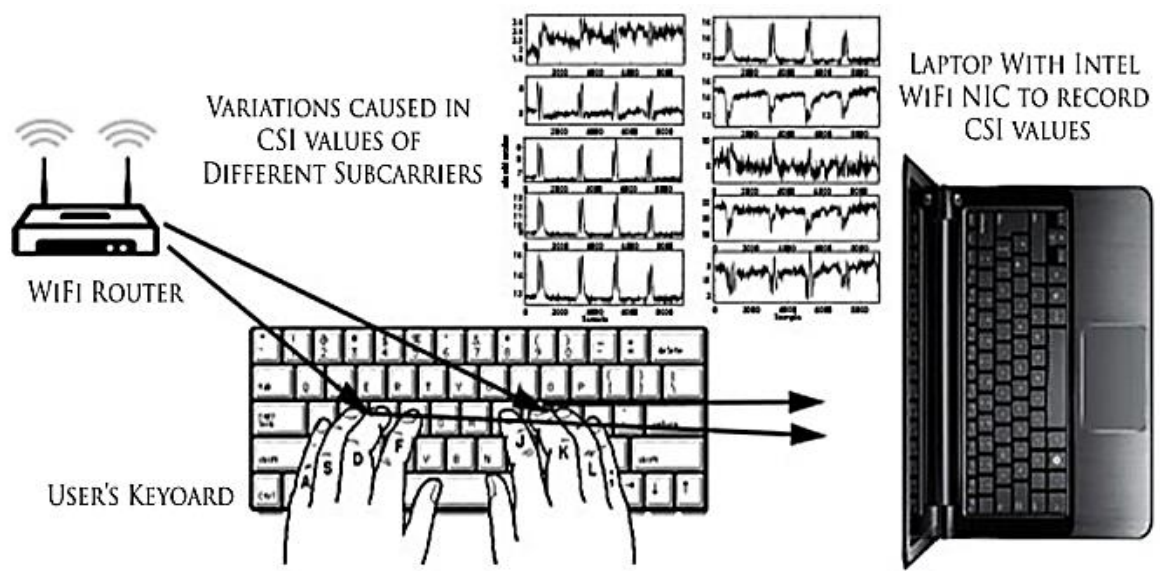

Fig. 3. Authentication mechanism using keystroke analysis.

### *E. Animal Detection*

Ref. [5] Nidhi Daxini, Sachin Sharma and Rahul Patel discovered that real-time animal detection system will reduce animal intrusion. These are possible using Viola and one algorithm for facial element detection. The video is taken from a camera and is changed over into surrounding mount. Identifying various pictures, the database has a Positive and negative terminal. Positive pictures have detected animals and negative pictures have non-detected pictures. Here component extraction strategy is used and later testing the classifier checks the program. Even this may create the wrong outcome if extraction isn't done and in the event, we don't have a tremendous training set.

### *F. Human Intrusion Detection*

In the approach, [6] the author works with human intrusion detection system using IR sensor, Wi-Fi module, Arduino, IOT, Atmega328p, and Temperature sensor. The Arduino Uno is used as the center of the system. It gets input at whatever point the movement is identified through PIR sensors. This system is acknowledged using PIR sensors [Fig. 4] to identify human existence MQ135 gas and smoke identifier, LM35 temperature sensor. The system is associated with the web using Wi-Fi module for example ESP-12e. To interface with the web and send parameters to the IOT stage we are using HTTP requests. To make alarms www.ubidots.com gives trigger event service.

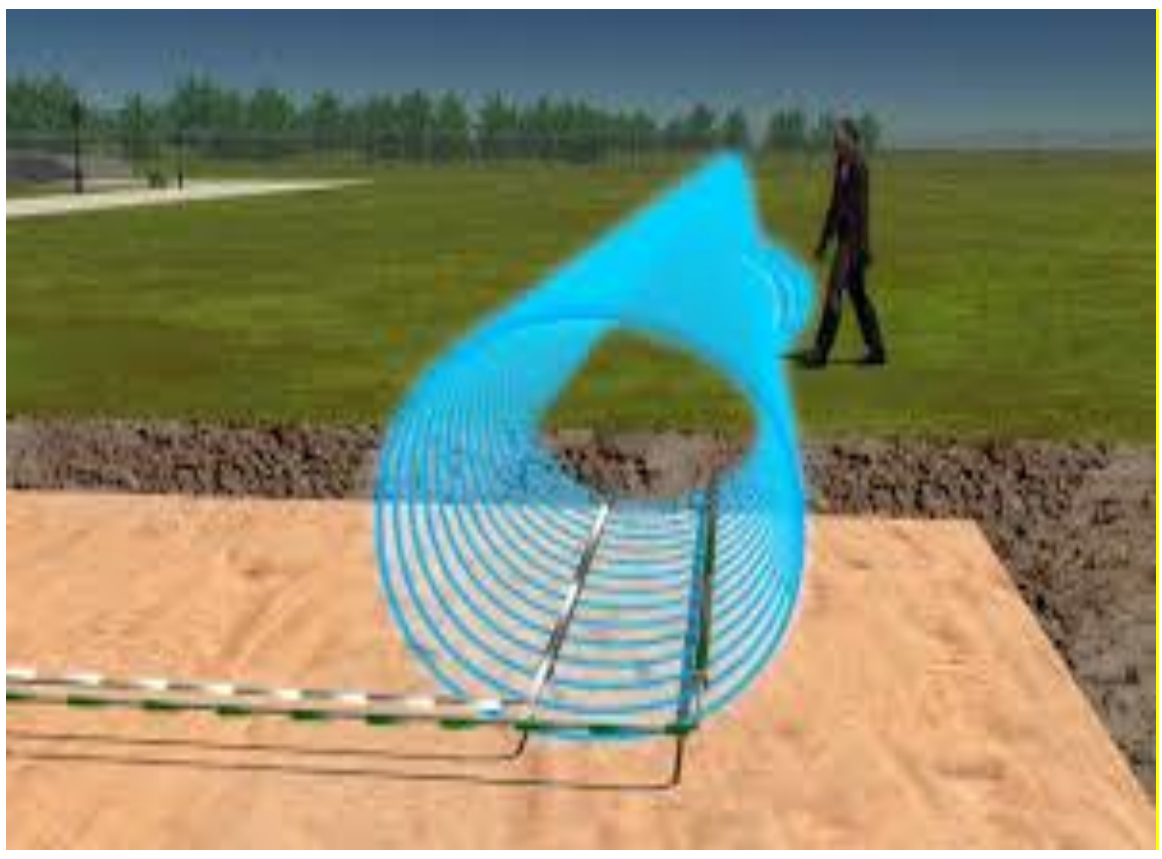

Fig. 4. System is using PIR sensors to detect human presence.

### *G. RFID*

RFID [7], Radio Frequency Identification is a reasonable innovation, can be Radio Frequency Identification is a reasonable innovation, can be executed in a few applications, for example, security, resource tracking, individuals tracking and get to control applications. The main purpose of this paper is to structure and actualize a digital security system which can convey in a secured zone where just a genuine person can be entered. We executed a security system containing entryway locking system using passive kind of RFID which can enact, confirm, and approve the user and open the entryway continuously for secure access. The benefit of using passive RFID is that its capacities without a battery and passive labels are lighter and are more affordable than the active labels. A unified system deals with the controlling, exchange and activity task. The entryway securing system works ongoing as the entryway opens immediately when the user put their tag in contact with the reader. The system additionally makes a log containing registration of every user alongside fundamental data of the user.

### *H. Parking System*

In the proposal, [8] the creator work with car parking system using Node MCU, IR sensor, LED. This system helps in arranging the parking slot and encourages the driver to achieve their parking spaces effectively as they realized which space is empty. The parking spot can be recognized using an Infrared sensor that interfaces with the ESP12-E (Node MCU) module that was programmed through the Arduino IDE. Users can get to parking spot data using a cell phone by means of an application. Particularly for users who have been enlisted previously, they have a code for login the application as the necessity for security system and user parking suitable [Fig. 5]. The system can work with the goal of research properly. In the system, the IR sensor is used for identifying the parking spot. The IR sensor is attached with the microcontroller Node MCU. Node MCU acts like a middle road between the sensors and the cloud. The Node MCU then transmits this information to the Firebase through Arduino IDE. The mobile application acts as an interface for the users to collaborate with the system on Cloud Firebase. This publication is helping us to work with IR sensor and Node MCU, firebase and the cloud server associated with the web it may be controlled remotely from anyplace on the world.

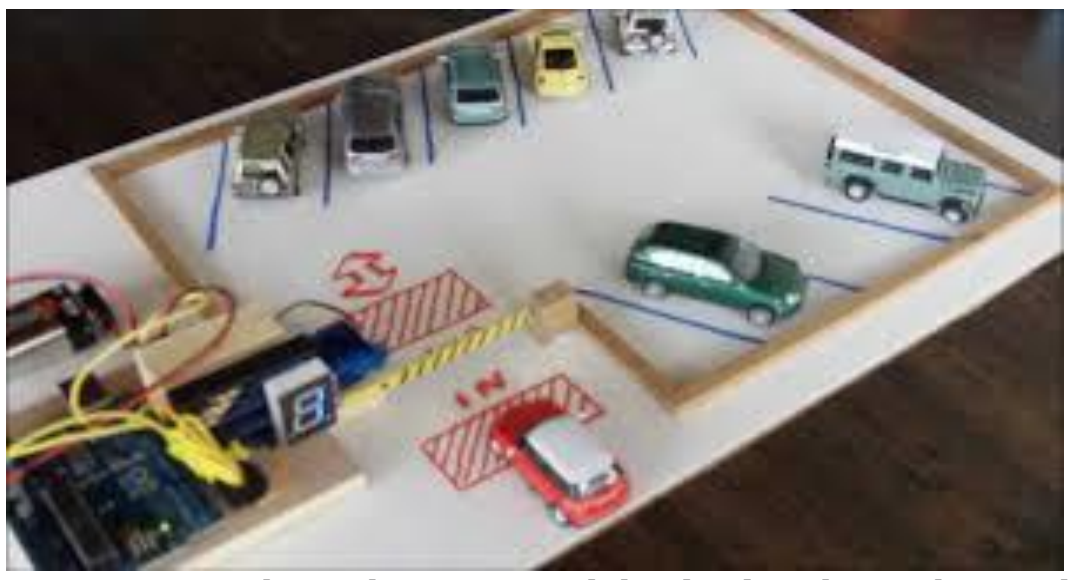

Fig. 5. Car entering the parking area and the display shows the number of vacancies in the parking slot.

This publication is related to the projects that can help us to work with IR sensor and Node MCU and applications that work with the cloud server.

These journals are related to our projects. It helped us to build the system and gave support to our theoretical knowledge so that we could work with IR sensor and Node MCU. Moreover, we tried to learn more about different apps that bridges between with cloud server and the Node MCU. These topics gave us an abstract idea that how actually a system can be implemented for border system, security purposes, using Infrared sensor in parking systems and ultrasonic sensors for distance calculation etc.

## III. Methodology

### *A. Theorem 1*

Here we have come through a plan diagram to get rid of the problem. Our basic idea is illustrated over the system flow diagram in Fig. 6. Here the diagram clearly states the flow of the system. Initially in the system the IR sensor detects the change and gives the signal to Node MCU. When the Node MCU receives the signal it sends to the BLYNK cloud and over there the data is processed and passed to the android app. From the app the notification is received and the buzzer and light is turned on.

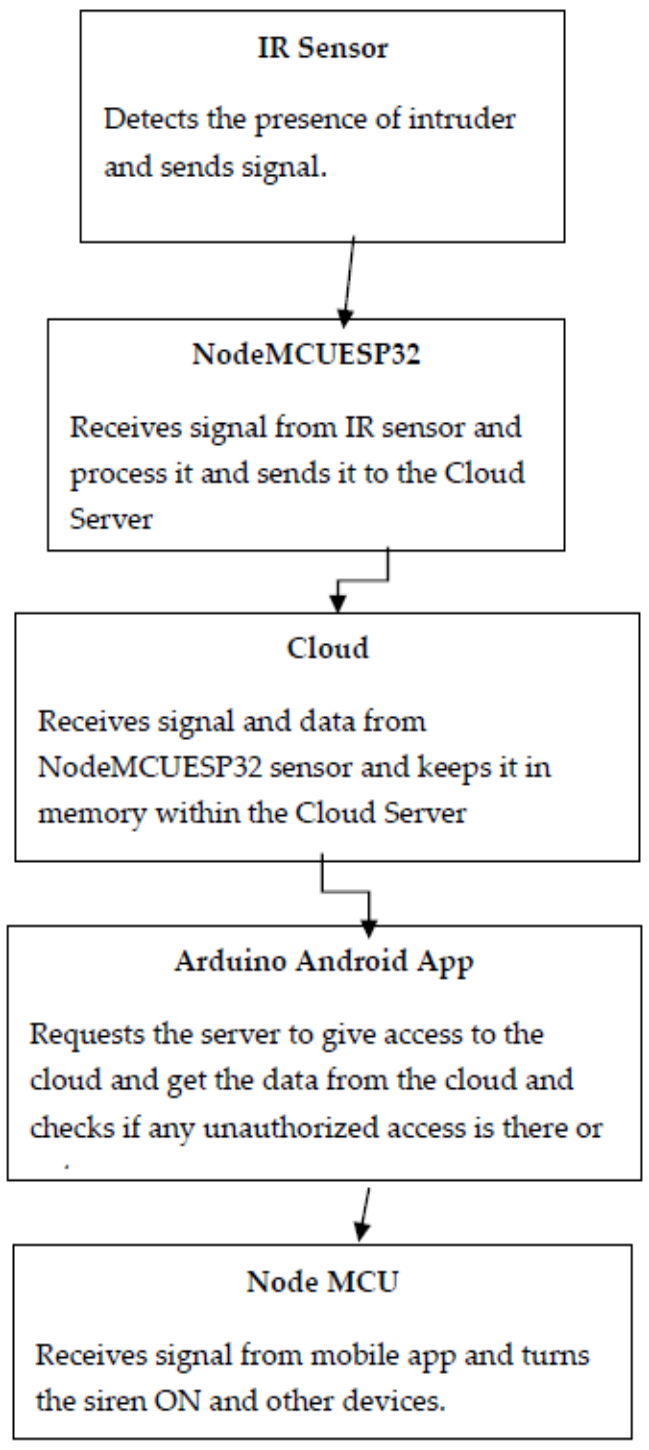

Fig. 6. System flow diagram.

### B. *Methodology Description and Its Functions*

For building up the project of the system proposed, we are working with Node MCU esp32. Node MCU is an open source IoT stage. It includes firmware which keeps running on the ESP32 Wi-Fi SoC from Espressif Systems and equipment which depends on the ESP-12 module. Node MCU esp32 Wi-Fi module is working with an infrared sensor (IR). The location scope of the IR sensor is around 2 cm to 30 cm. The sensor module is interfaced with Arduino having IO voltage level of 3.3V to 5V. Most Importantly, if an object detects by the IR sensor and after that the signal pass to the Node MCU esp8266 Wi-Fi module. Node M CU will check whether the Wi-Fi is connected or not. From that point forward, Node MCU esp8266 Wi-Fi module will check the signal from the IR sensor. And after that, the Node MCU esp8266 Wi-Fi module analysis the signal through Arduino IDE and send an HTTP request to the BLYNK cloud server. At the point when the user needs to see the touch history, then the web sends a request to the cloud server. Cloud server synchronizes the data from the database and response to the web server. At long last, the user can see details of the object which cross the IR sensor.

Let's say when an intruder arrives near the border and finds out that there is no soldier near the border or no security is there. He or she is trying to enter the border area illegally by taking the advantage of the current condition. The intruder steps in and blocks the signal transmission of the device. Then immediately alarm notification will be sent to the nearby army soldier. The soldier will be notified and will move towards the place where the intruder is. And can warn the intruder and take proper action that is needed to be done for breaking the law. Symbolic representation is shown in Fig. 7.

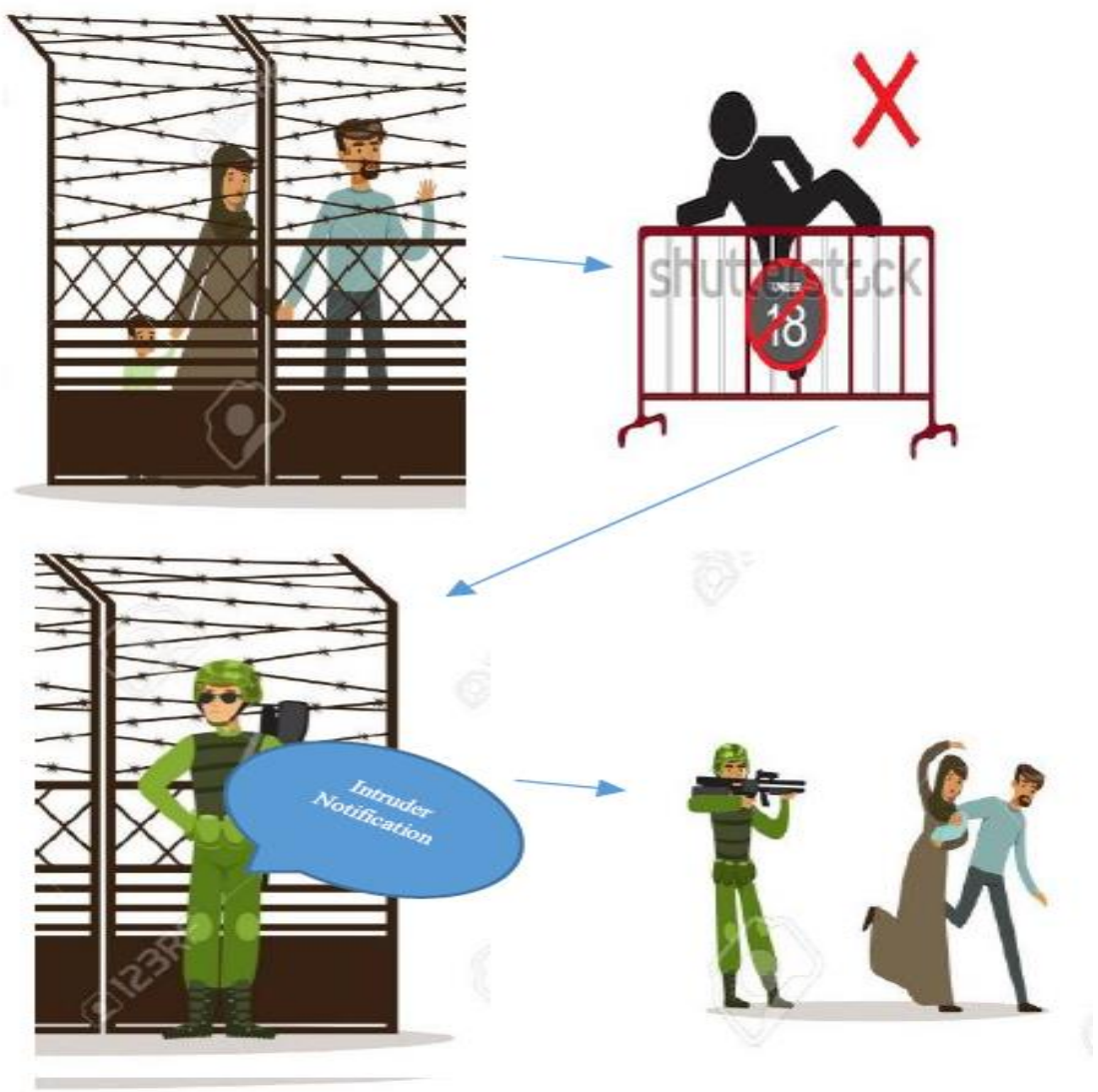

Fig. 7. Expected method of working during the intruder visit and steps to be taken.

### A. *Layout of the System*

To make this project we use IR sensor, node MCU, BLYNK cloud server, Node MCU always connected to the server (cloud) by Wi-Fi service and have an ability to send the HTTP request to the server by giving a notification. This notification stores in the server database and server notice center update. When server update, then this update notification sends to the user. User has an Arduino android app (Blynk app), and Web quickly observes this notification. Then from the app alarm is turned ON. Individual devices are shown in Fig. 8.

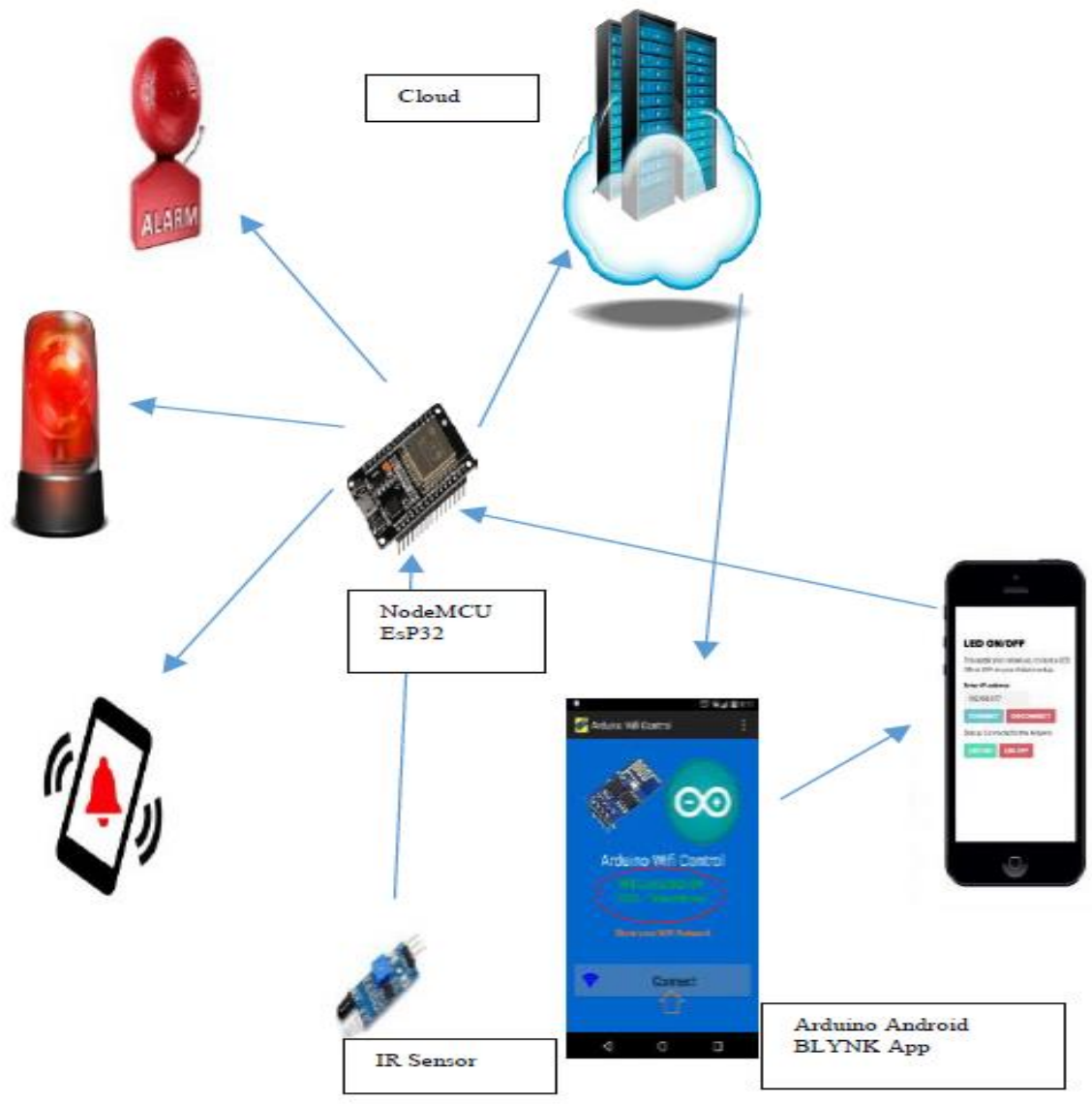

Fig. 8. Overall system function.

### B. *Discussion*

We have implemented a simple basic circuit diagram to connect the system with the send signal via Wi-Fi. There are two following circuits. One of it connects D1 pin of the NodeMCU with the GND pin of the Buzzer and LED light represented in Fig. 9. And the other circuit shows how the infrared sensor is connected to the NoeMCU by connecting the wires with the source voltage of 3V, GND and analog pin (A0) [Fig. 10]. The basic diagram is given below:

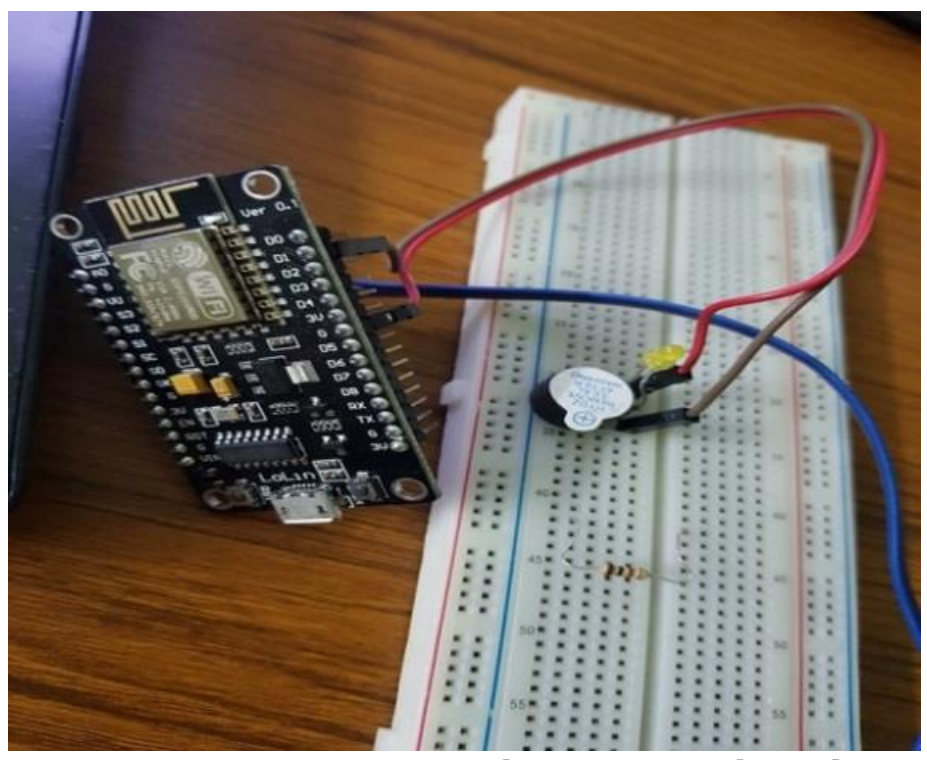

Fig. 9. Circuit connection, connecting the D1 pin with NodeMCU and the GND pin with buzzer and the LED light.

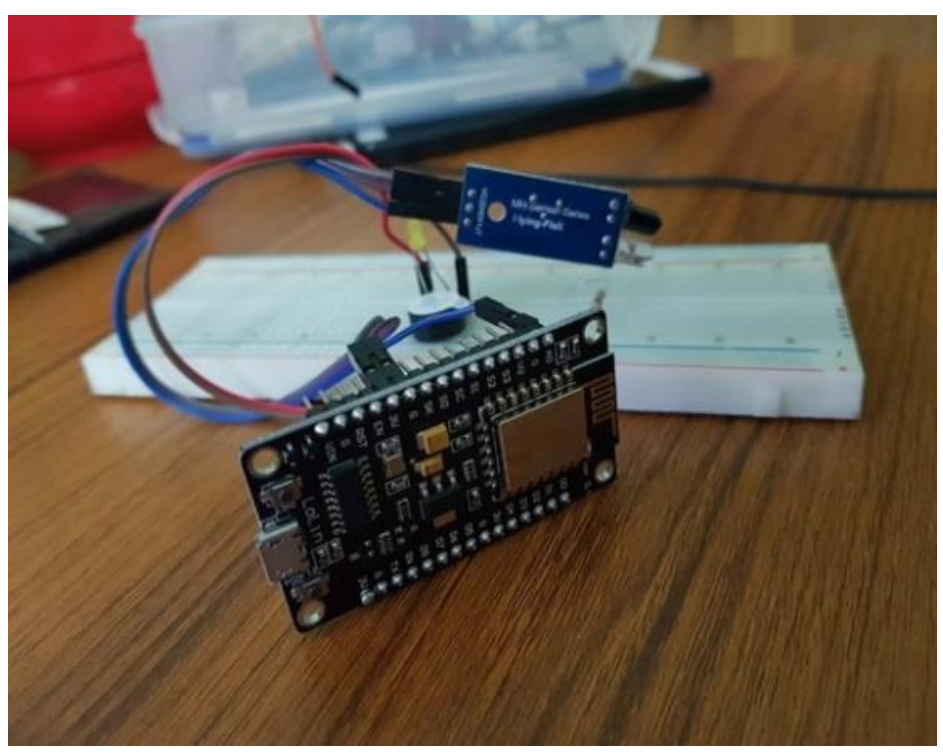

Fig. 10. Circuit of NodeMCU connecting the 3v, GND and A0 with the IR sensor.

When an intruder crosses the respective area, the signal is sent to Node MCU. It is directly sent via Wi-Fi to the Blynk cloud. It then sends a notification to the app (smartphone). See Fig. 11 the notification is taken from the app. Inside the app, we can see the notification message. And then we can go to the BLYNK app interface and from there we can turn ON the buzzer from the app. In Fig. 12 we can see the Buzzer is currently OFF. When the Buzzer is pressed it will turn ON the Buzzer. In the future, we will add a camera on this project and we will improve our detection area as well as we can use the ultrasonic sensor instead of the IR sensor. It can be used to detect a long-range object. For better and high-speed notification system we can use raspberry pi 3 B+ for make server request. Web and Android app user interface design can make more user-friendly and add more feature better user experience. We can use better and advance database system to add more data and space.

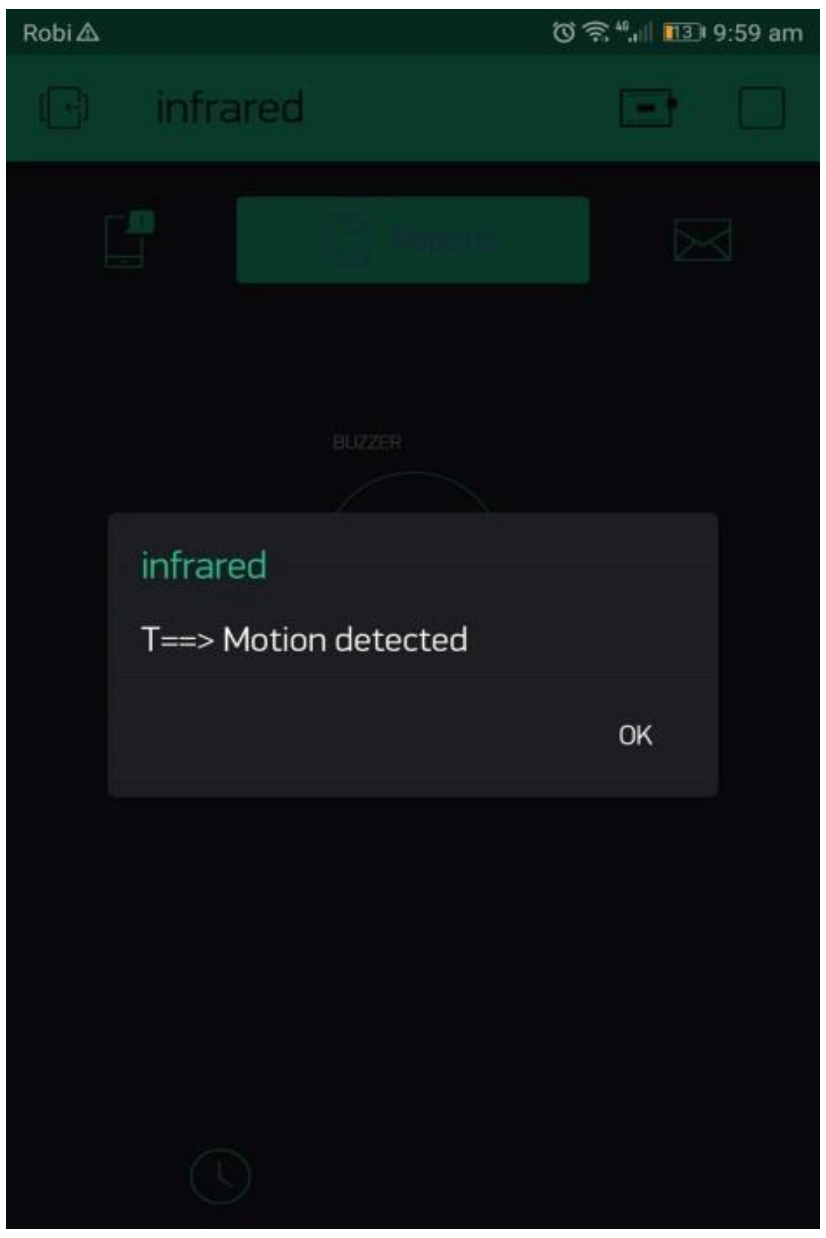

Fig. 11. Motion detected and the notification message pops up over the smartphone.

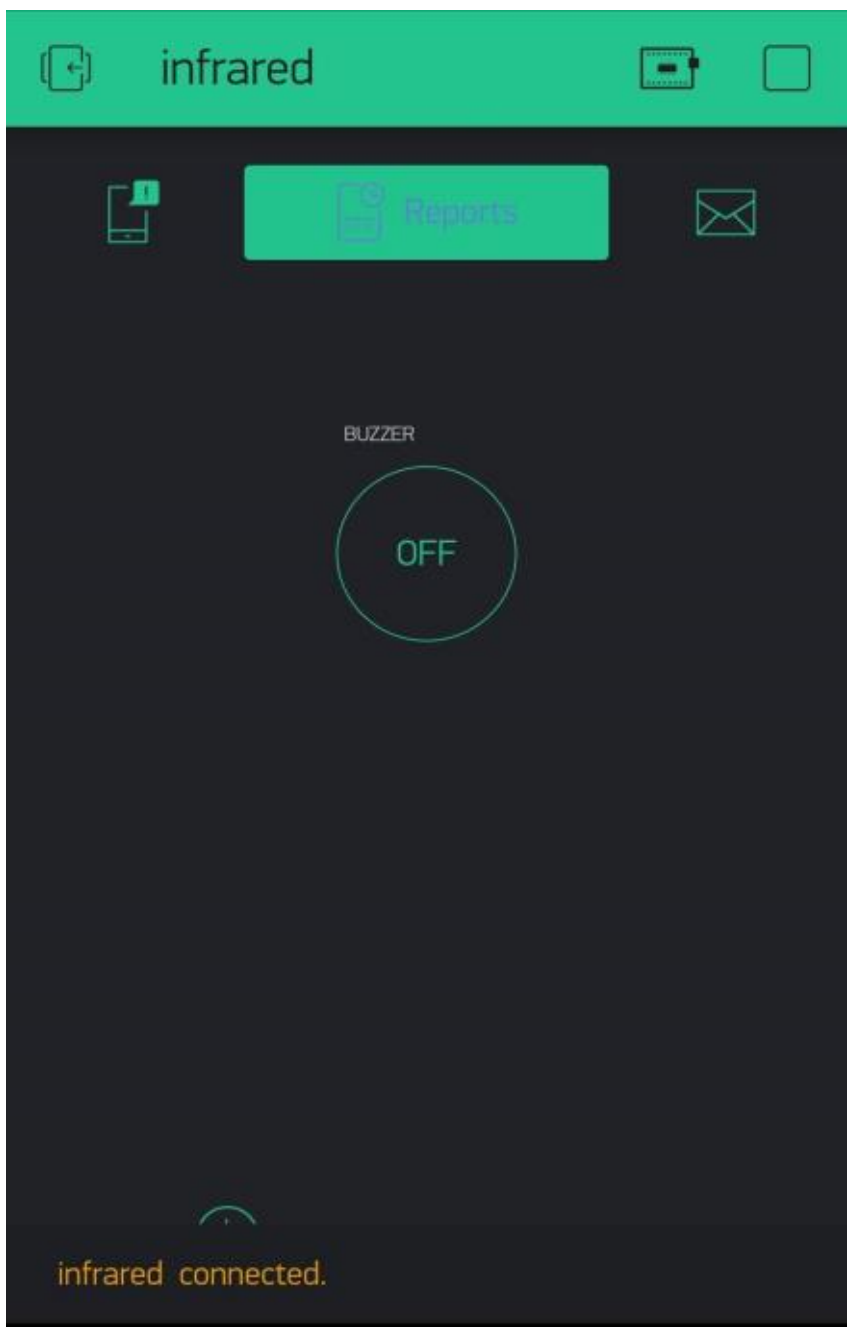

Fig. 12. Blynk app interface where the buzzer alarm is currently OFF. When it is turned ON the Buzzer and lights is turned ON.

## IV. Materials and Methods

### A. Hardware Device

- Node MCU esp32:
- The Node MCU ESP-32S [9] is one of the advancement board made by Node MCU to assess the ESP-WROOM-32 module. It depends on the ESP32 microcontroller that flaunts Wi-Fi, Bluetooth, Ethernet, and Low Power bolster all in a solitary chip. Propelled API for equipment IO, which can significantly decrease the excess work for designing and controlling equipment. Code like Arduino, yet intelligently in content. Occasion driven API for system applications, which facilitates designers composing code running on a 5mm*5mm estimated MCU in Nodes style. Significantly accelerate your IOT application creating process. Under WI-FI MCU ESP8266 incorporated and easy to prototyping improvement pack. We give the best stage to IOT application advancement at the most minimal expense.
- BLYNK app
- Wire
- Breadboard
- IR Sensor
- An infrared sensor [10] is an electronic gadget with simple configuration, that radiates so that we can detect a few parts of the environment. An IR sensor can quantify the warmth of an item just as it recognizes the motion when an object is passed in front of it. These sorts of sensors emit infrared radiation and receive the radiation via the receiver that is called passive Infrared sensor.
- Buzzer
- LED

In our task, we are working with Node MCU esp32. Node MCU is an open source IoT stage. It incorporates firmware which keeps running on the ESP32 Wi-Fi SoC from Espressif Systems and equipment which depends on the ESP-12 module. [10] Node MCU esp32 Wi-Fi module is working with an infrared sensor (IR). The recognition scope of the IR sensor is nearly around 2 cm to 30 cm. The sensor module is interfaced with Arduino having IO voltage dimension of 3.3V to 5V.

The IR sensor has IR Transmitter that transmits the rays and IR receiver which receives the ray after getting any obstacle. It is shown in block diagram in Fig. 13.

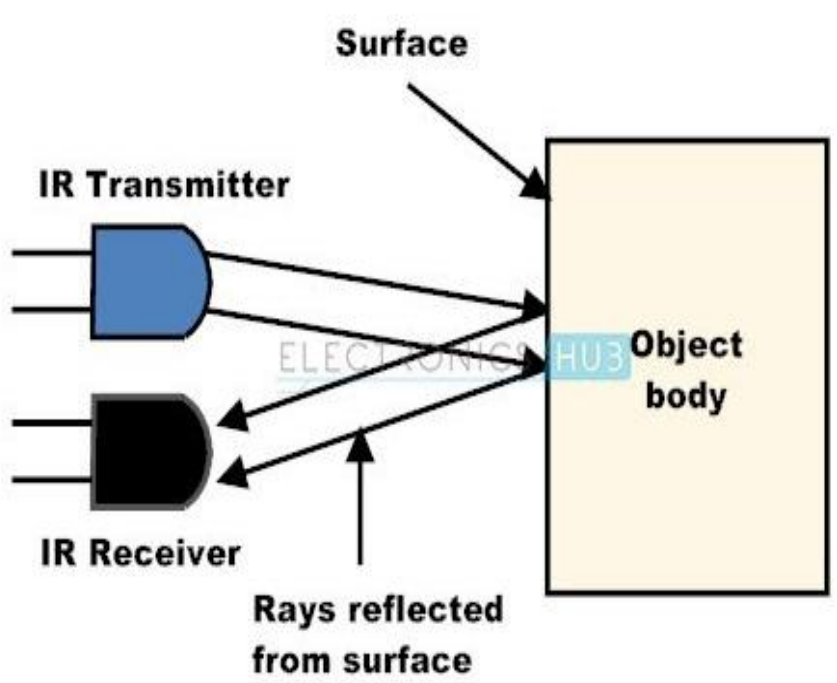

Fig. 13. IR transmitter and IR receiver sending and receiving rays. [10]

An IR sensor, **A** is mounted over a wall there is a gap between the human and the IR is distance **B**. The rays split accordingly that when the human enters that range the rays will get the obstacle and get reflected earlier. It is represented in Fig. 14.

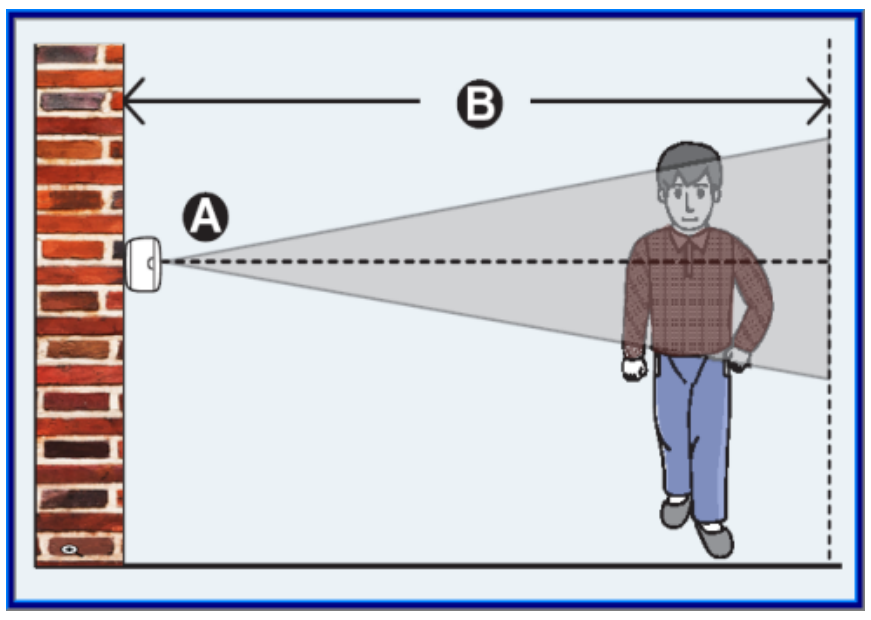

Fig. 14. Representing human intervention with diagram how IR transmitter and IR receiver works.

As a matter of first importance, if an item distinguishes by the IR sensor and after that, the flag goes to the Node MCU esp8266 Wi-Fi module to the web server from where the HTTP request is send to the NodeMCU. Node MCU will check whether the Wi-Fi is associated or not. From that point onward, Node MCU esp8266 Wi-Fi module will check the flag from the IR sensor. [11] And after that, the Node MCU esp8266 Wi-Fi module investigation the flag through Arduino IDE and send an HTTP (response) solicitation to the BLINK cloud server. This is shown in Fig. 15.

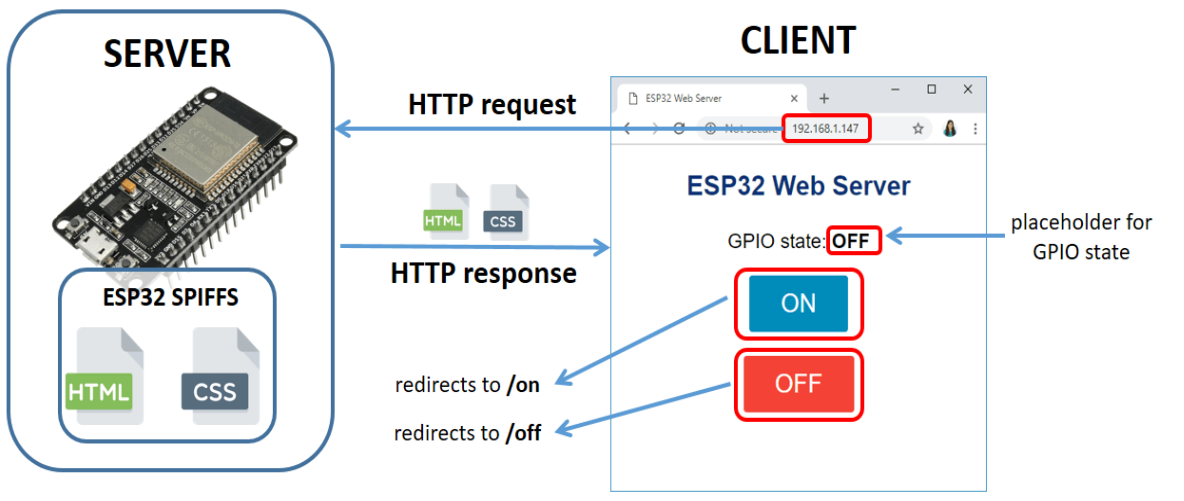

Fig. 15. [11] Determining how Http request is sent over internet to the BLYNK cloud and its app.

When a user needs to see the touch history, then the web sends a request to the cloud server. Cloud server synchronizes the data from the database as the data is categorized in NFS and CIFS and is stored in the google cloud. But our is stored int the Blynk cloud and response to the web server. [12] Finally, the user can see details of the time and date of the intrusion. This are handled in data centers. Represented in Fig. 16.

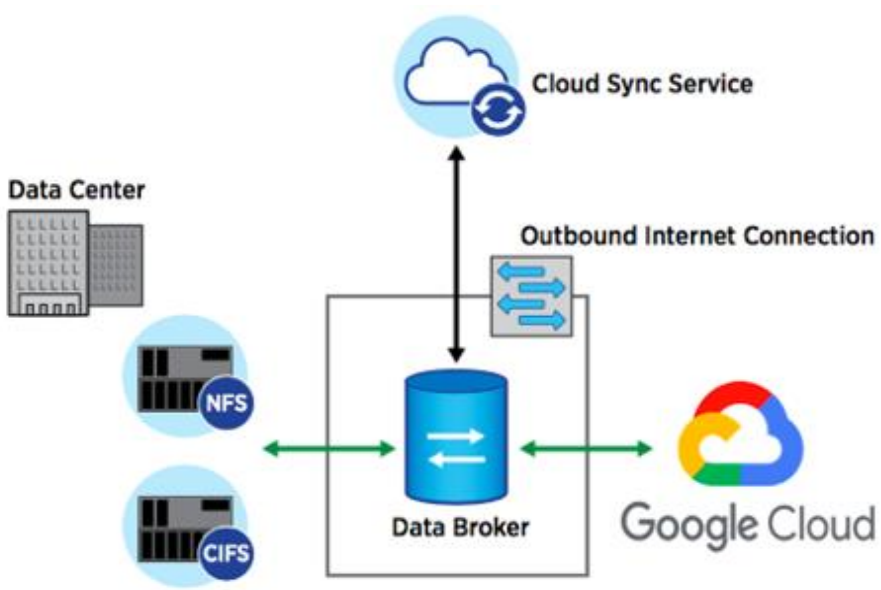

Fig. 16. Ref. [12] How data is stored in google cloud and synced.

Over this project we use IR sensor, Node MCU, BLYNK cloud server, Node MCU always connected to server by Wi-Fi service and have an ability to send http request to server when a new Follower (Intruder) is found it goes to the database to check what message is written by the admin to give to the user. Using functions and algorithm it is forwarded to FCM and from the FCM (cloud) the message as notification goes to the user. This is shown in Fig. 17.

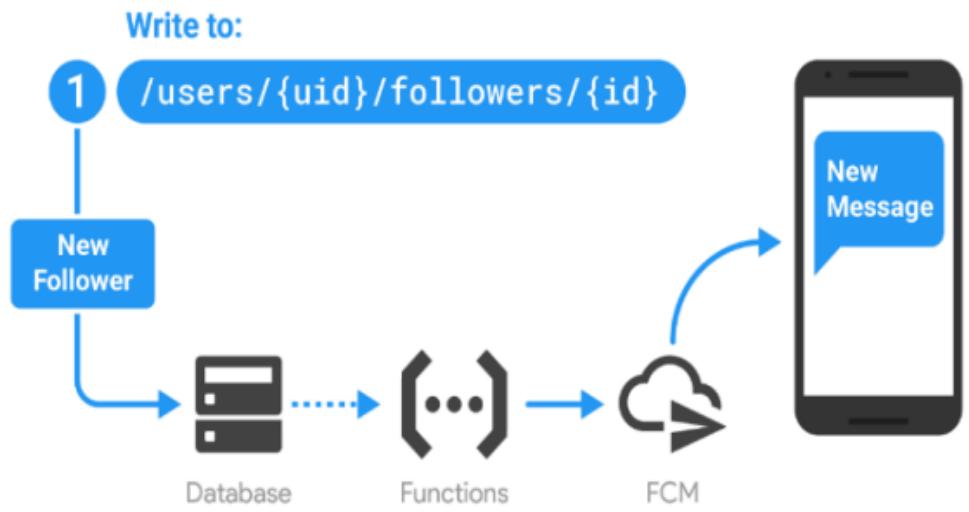

Fig. 17. New message is sent to the user when new follower arrives. And the path it follows to reach the user.

This notification stores in server database and server notice center update. [13] When server update, then this update notification sends to user. User have an Arduino android app, and web immediately see this notification. The notification is given by sending a particular message over the Blynk app. Which ensure the user about the intruder that entered the restricted area.

Initially, we included the libraries and used the authentication for the app then took the input from the user and the partial execution code of the whole system is being shown Fig. 18.

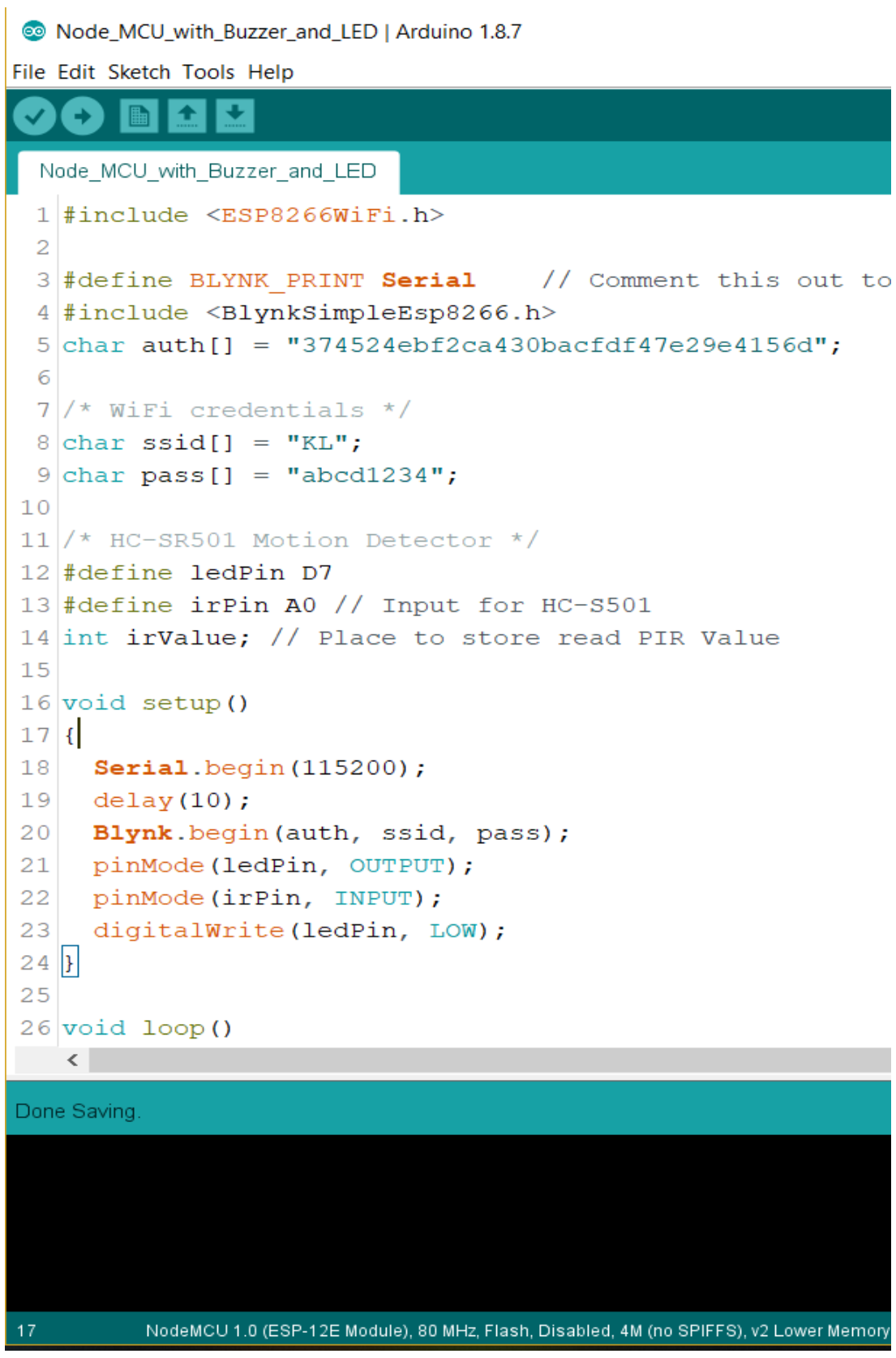

Fig. 18. Execution code for the system (1).

Below is the code of the system that is used for the system to execute the process that has been discussed earlier. Here the range is compared within which if the intruder enters it will compare the value and execute its instructions. In Fig. 19 is the execution code of the process is shown.

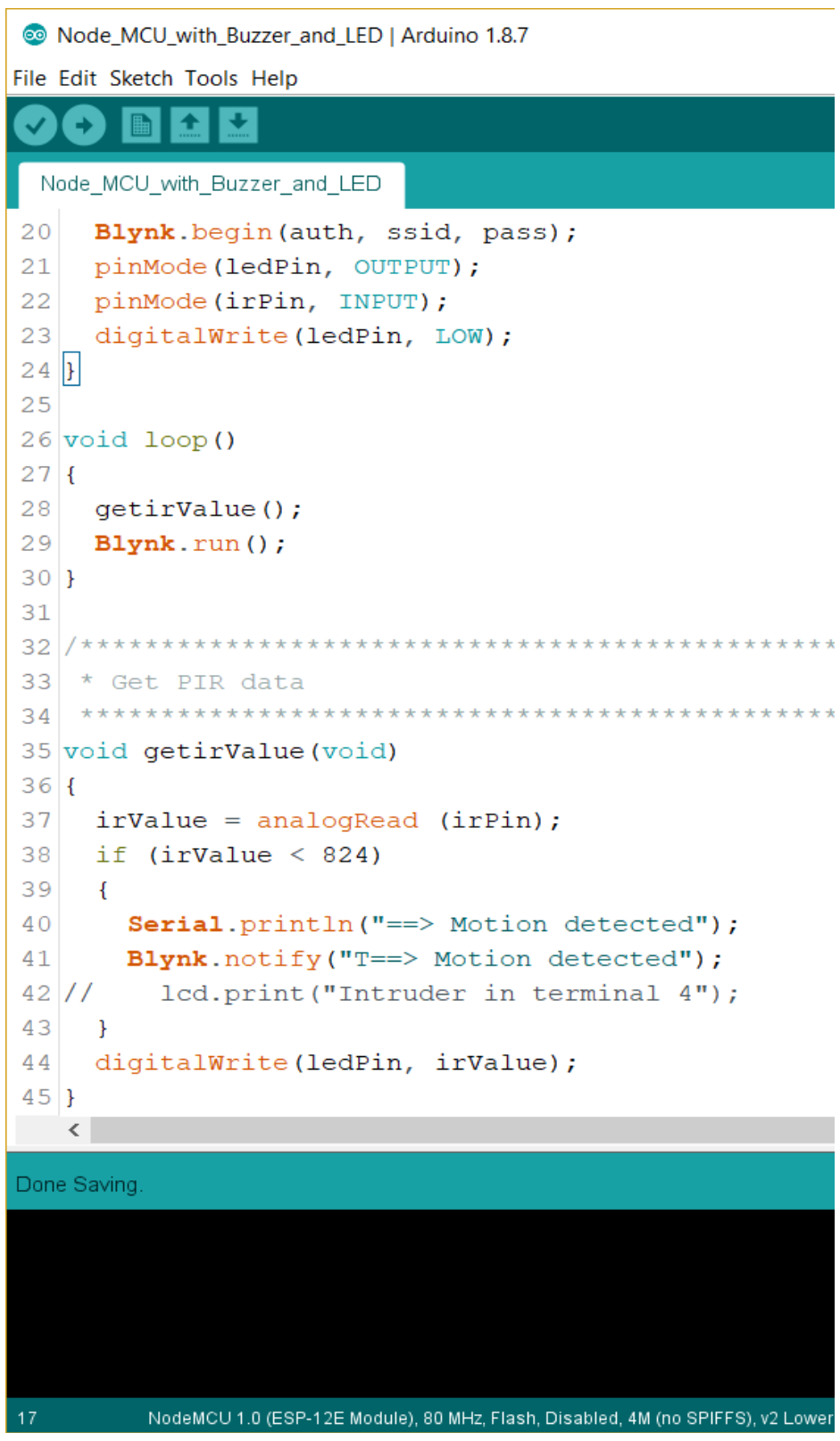

Fig. 19. Execution code for the system (2).

The System is working properly after the execution of code and setting up the system properly using the LED, Buzzer and Node MCU.

Here, the device gives the notification of the intruder and the alarm is set ON, we can see it in the Fig. 20 below.

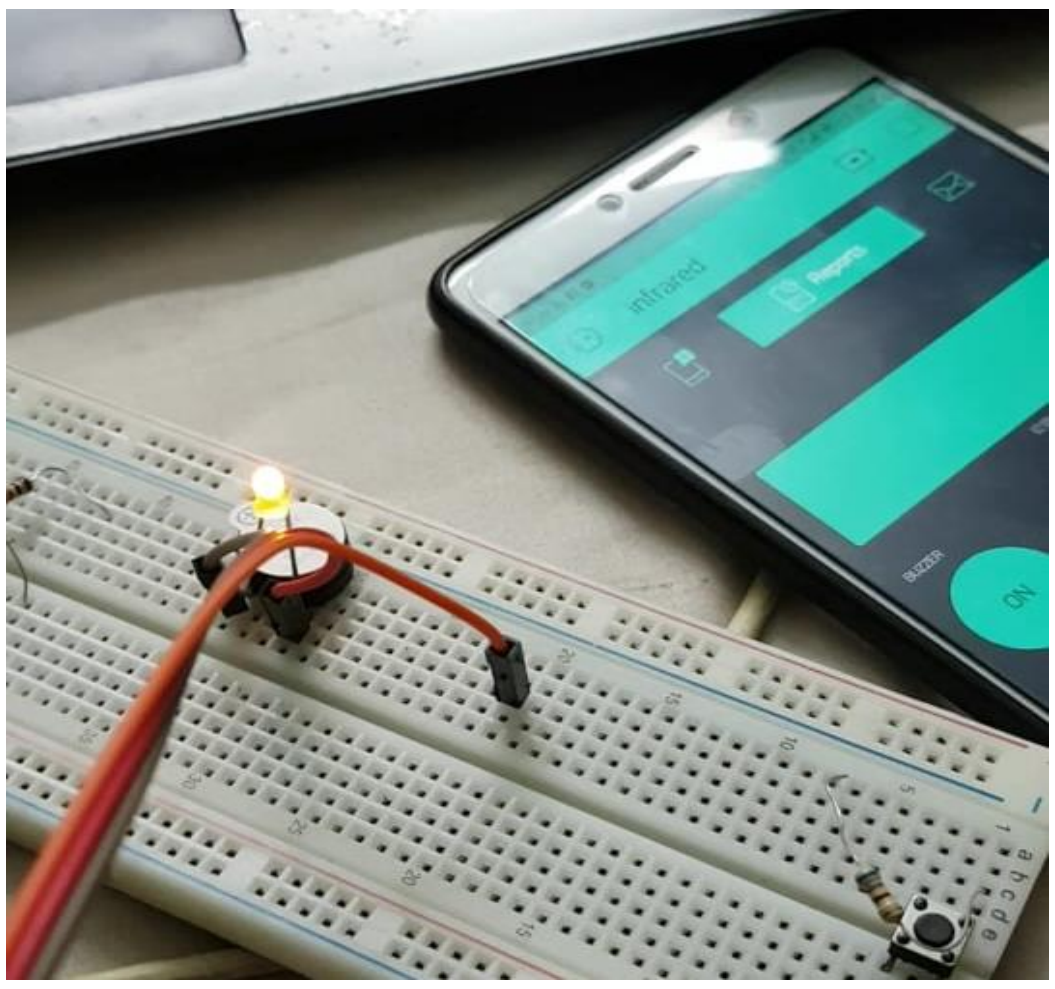
Fig. 20. Working of the project. When the switch is turned on from the Blynk.

## V. Conclusion

We have seen how much hard work the army and the soldiers over the border do to keep their countries safe from terrorist attack and illegal people of other countries. It is really difficult to stay 24 hours fully active for the soldiers. To reduce their effort of the soldiers the system can be implemented and the security of the nation can be ensured. This system is really cost efficient and very easily usable for the soldiers.

## Conflict of Interest

Afsana Tasnim had an opinion of using Raspberry PI for the server connection. But the project would be expensive to use. So, Shawan Shurid decided to work it with Node MCU to reduce cost and make it easier. If we could use Raspberry PI, we would be able to get more features but the cost would increase 7 times more. Other than that, there was no conflict while doing the project.

## Author Contributions

Shawan Shurid and Afsana Tasnim together conducted the research and system specification. Shawan Shurid analyzed the sytem architecture and built the circuit and coded for establishing connection with server. Afsana Tasnim wrote the paper and managed the connection between the user application (BLYNK App). AKM Bahalul Haque was responsible for overall idea generation and supervision of the project. Reading, reviewing the project and working on it.

## Acknowledgement

We respect and thank our honorable faculty AKM Bahalul Haque, for providing us an opportunity to do the project work in detection of illegal border cross and giving us all support and guidance, which made us complete the project properly. We are extremely thankful to him for providing such a nice support and guidance, although he had busy schedule managing the corporate affairs. We were really great full and blessed to work under his guidance.

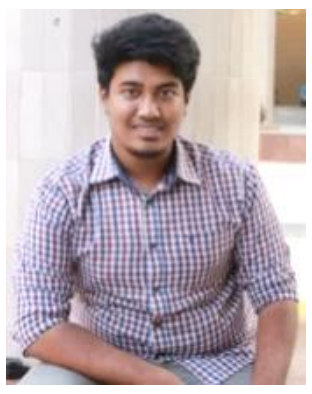

**Shawan Shurid** is currently a student of North South University studying computer science engineering where he is going to complete his under graduation by 2020. His majors are computer networking, computer security, web developer (word press, PHP developer). He also has expert knowledge on C, C++ , Java, Python and Assembly Language. Partially he works with IOT devices for social security and other purposes.

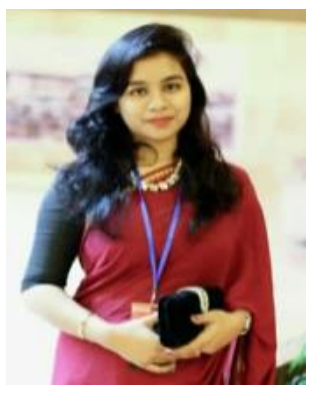

**Afsana Tasnim** is currently a final year student of North South University pursuing BS in computer science engineering and will be graduating in 2020. She is an ardent programmer in software engineering and programming languages like C, C++, Java and net technologies. She has a strong interest in the field of computer security and information assurance. Her research interest includes areas of computer security, operating systems and mobile security.

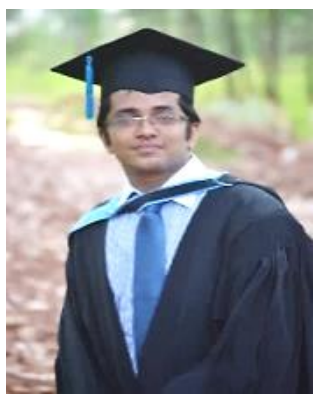

**AKM Bahalul Haque** is currently working as a faculty member of Department of Electrical and Computer Engineering in North South University. He completed his bachelor in computer science from Bangladesh and his master of science in information technology from Germany. His area of research is cyber security and software engineering. He has research papers published in International Journal and Prestigious conferences.